\documentclass[twocolumn]{aastex631}

\usepackage[version=4]{mhchem}
\usepackage{comment}
\usepackage[T1]{fontenc}
\usepackage{balance}
\usepackage{hyperref}

\shorttitle{Constraining exoplanet interiors using observations of their atmospheres}
\shortauthors{Lichtenberg, Shorttle, Teske, Kempton}

\graphicspath{{./}{figures/}}

\begin{document}

\title{Constraining exoplanet interiors using observations of their atmospheres}
\correspondingauthor{Tim Lichtenberg}
\email{tim.lichtenberg@rug.nl}

\author[0000-0002-3286-7683]{Tim Lichtenberg}
\affiliation{Kapteyn Astronomical Institute, University of Groningen, Groningen, Netherlands}

\author[0000-0002-8713-1446]{Oliver Shorttle}
\affiliation{Department of Earth Sciences, University of Cambridge, Cambridge, UK}
\affiliation{Institute of Astronomy, University of Cambridge, Cambridge, UK}

\author[0009-0008-2801-5040]{Johanna Teske}
\affiliation{Earth and Planets Laboratory, Carnegie Institution for Science, Washington, DC, USA}
\affiliation{The Observatories of the Carnegie Institution for Science, Pasadena, CA, USA}

\author[0000-0002-1337-9051]{Eliza M.-R. Kempton}
\affiliation{Department of Astronomy and Astrophysics, University of Chicago, Chicago, IL, USA}
\affiliation{Department of Astronomy, University of Maryland, College Park, MD, USA}

\begin{abstract}

Astronomical surveys have identified numerous exoplanets with bulk compositions that are unlike the planets of the Solar System, including rocky super-Earths and gas-enveloped sub-Neptunes. Observing the atmospheres of these objects provides information on the geological processes that influence their climates and surfaces. In this Review, we summarize the current understanding of these planets, including insights into the interaction between the atmosphere and interior based on observations made with the James Webb Space Telescope (JWST). We describe the expected climatic and interior planetary regimes for planets with different density and stellar flux and how those regimes might be observationally distinguished. We also identify the observational, experimental, and theoretical innovations that will be required to characterize Earth-like exoplanets.

\end{abstract}

\section*{Review Summary} 

\textbf{Background:}
Extrasolar planets (exoplanets) orbit stars other than the Sun. Most of the $\approx$6000 known exoplanets were detected indirectly by studying how they affect the light emitted by their host stars. The principal observational signatures are the wobble that exoplanets induce in the star’s radial velocity and the periodic dimming of the star’s light as the exoplanet passes through our line of sight to their host star, which are known as transits. These techniques provide mass and radius measurements, respectively.

Observed exoplanets display a wider diversity than that of the planets in the Solar System. This includes their physical and chemical characteristics, orbital configurations, stellar irradiations, and compositions. Characterization of exoplanet properties is provided by astronomical observations. Interpreting those properties requires an understanding of planetary science, geophysics, geochemistry, atmospheric science, and astrobiology.

\textbf{Advances:}
Transit surveys by use of space telescopes have provided statistical insight into the population of exoplanets. Those surveys have found a common class of small exoplanets that is intermediate in size between Earth and Neptune and spans a large range in bulk densities, from close to Earth-like (super-Earths) to densities that necessitate large amounts of atmospheric volatiles (sub-Neptunes). These observations have allowed researchers to perform comparative planetology across a statistically meaningful sample of exoplanets.

\setcounter{figure}{-1}
\begin{figure*}[tbh]
\centering
\includegraphics[width=0.85\textwidth]{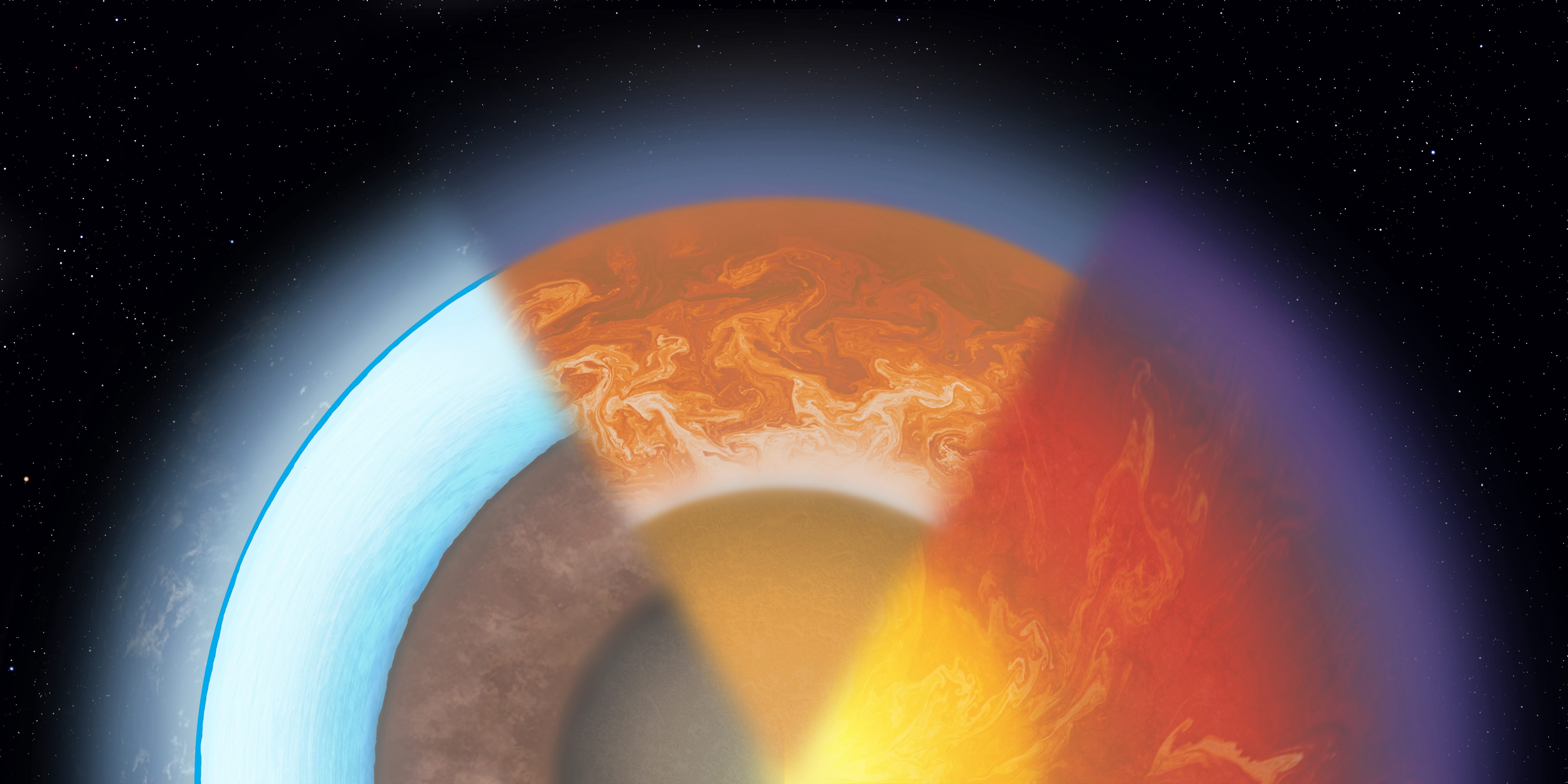}
\caption{\textbf{Predicted interior structures of sub-Neptune exoplanets.} Observed density and atmospheric constraints are interpreted in three scenarios, each with a gaseous envelope. (\textbf{Left}) Water worlds have a small metal core, rocky mantle, high-pressure ice layer, and possible liquid water ocean. (\textbf{Middle}) Gas dwarfs have a larger metal core and a magma ocean. (\textbf{Right}) Global supercritical regimes have no clear boundaries between layers. Image credit: Mark A. Garlick.}
\label{fig:summary}
\end{figure*}

Improved precision in transit and radial velocity observations has provided bulk density measurements for numerous exoplanets, which can be compared with interior structure and evolution models. Spectroscopic observations made with the JWST have provided some information on the atmospheric composition of low-mass exoplanets. These combined observations have shown a highly diverse population, including exoplanets composed of nearly pure iron, mixtures of rock and metal, and hydrogen-rich gas envelopes.

Most transiting exoplanets receive far more extreme irradiation from their host star than Solar System planets receive from the Sun. The resulting higher temperatures cause more material and energy exchange throughout the planet, blurring the distinction between segregated layers such as the atmosphere and rocky mantle. This interior-atmosphere exchange potentially allows the interior composition of a planet to be constrained by measuring the chemistry of its atmosphere. This indirect method could be used to study planetary geodynamics.

JWST spectra of several sub-Neptune exoplanets show evidence of chemical mixing between the upper atmospheric layers and the deep interiors. Spectroscopic observations of Earth-sized exoplanets orbiting low-mass stars show that they do not have thick atmospheric envelopes. However, several ultrashort-period super-Earths, with orbital timescales of <1 day, defy expectations of complete atmospheric evaporation: Their bulk densities necessitate either chemically undifferentiated interior structures or a high bulk abundance of volatile elements. In either case, this constrains the nature of interior-atmosphere exchange on these exoplanets.

\textbf{Outlook:}
Interpreting these results requires collaboration between researchers in exoplanet astronomy with those in planetary science, geophysics, and geochemistry. The limited astronomical data needs to be understood within laboratory-validated limits of material properties and thermodynamics. This approach will likely improve our understanding of how atmospheric escape, natal H/He envelopes, volatile accretion, and global-scale planetary magma oceans all interact to produce the properties of low-mass, highly irradiated exoplanets.

Upcoming facilities will perform wider and more detailed surveys to study planetary demographics across a larger temperature range. Future space telescopes are being designed to directly image temperate terrestrial exoplanets around Sun-like stars. This would provide the capability to investigate Earth-like exoplanets and assess their habitability.

\section*{Introduction}
\noindent

Observations of extrasolar planets (exoplanets) -- planets orbiting stars other than the Sun -- has increased our understanding of Earth and the Solar System \citep{1992Natur.355..145W,1995Natur.378..355M}. A great diversity of planetary mass, stellar irradiation, and planetary composition is evident among exoplanets, extending beyond the planets of the Solar System. Compared with exoplanetary systems, the Solar System architecture, with a Jupiter-like planet at a few astronomical units (au) distance from its host star, is unusual \citep{Fernandes2019,Zhu2021}, but it is unclear if Solar System-like terrestrial planets are equally rare \citep{Lichtenberg2025}.

\begin{figure*}[tbh]
\centering
\includegraphics[width=0.76\textwidth]{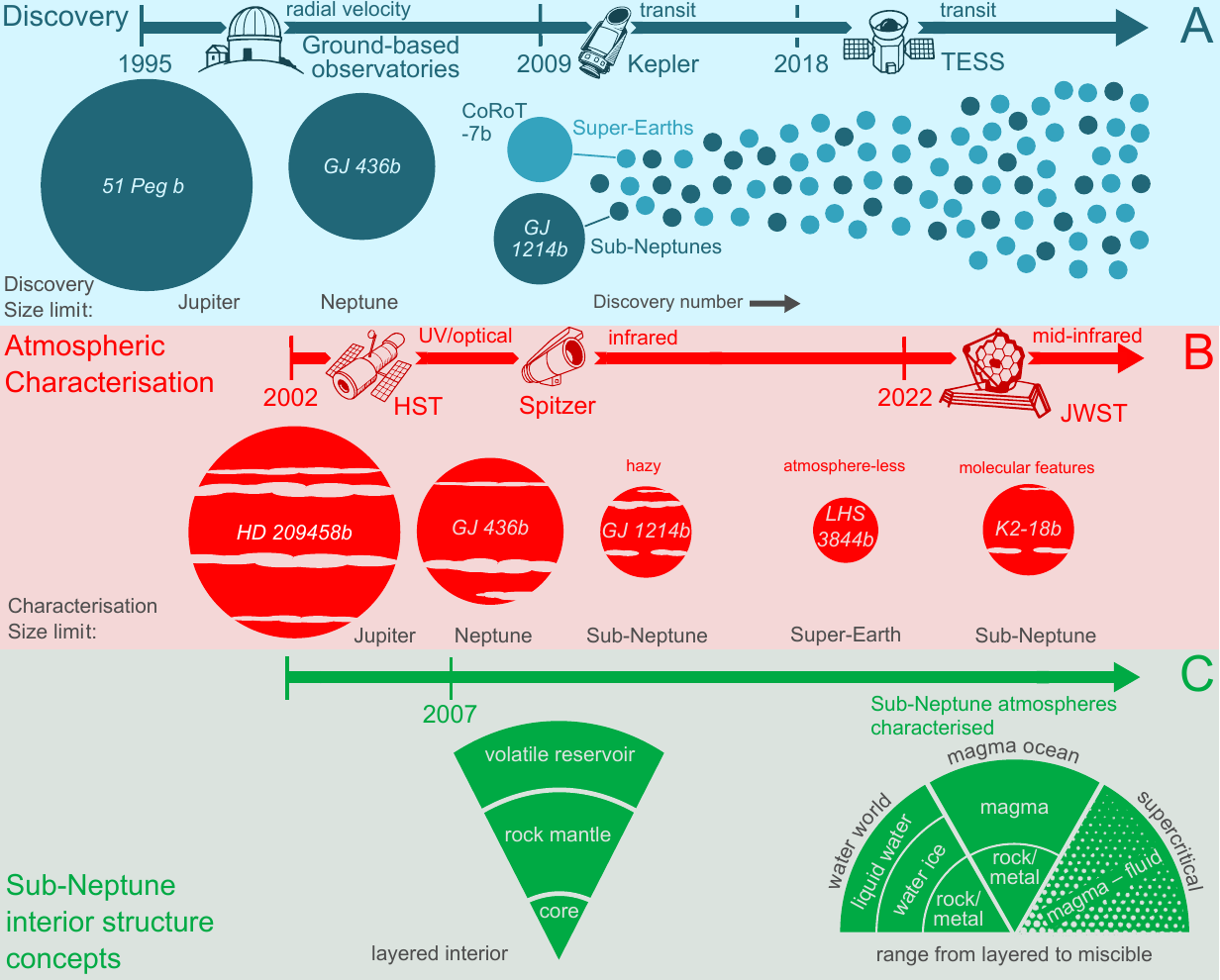}
\caption{\textbf{Timeline of exoplanet detection and atmospheric characterization.} (\textbf{A}) Symbol sizes schematically illustrate the sizes of exoplanets detected by each facility and the increasing numbers of super-Earths and sub-Neptune exoplanets found by using Kepler and TESS. (\textbf{B}) An equivalent illustration of the atmospheric characterization of exoplanets by use of space telescopes. (\textbf{C}) The conceptual interior structures of sub-Neptune exoplanets, with increasing complexity owing to improved characterization.}
\label{fig:many_worlds}
\end{figure*}

Exoplanet observations, with both ground-based and space-based instruments, have advanced beyond simply detecting exoplanets to providing detailed characterization of statistically meaningful samples (Fig.~\ref{fig:many_worlds}). There are about 6000 known exoplanets, spanning wide ranges of radii, masses, orbital configurations, and host star types. The most commonly detected exoplanets are intermediate in size between Earth and Neptune, which have no Solar System counterpart \citep{Howard2012}. Such planets form two sub-populations \citep{fulton17,vaneylen18}: larger and less dense objects typically referred to as sub-Neptunes [$\approx$2 to 4 Earth radii ($R_{\rm{Earth}}$)], and smaller, denser planets known as super-Earths ($\lesssim 1.6$~$R_{\rm{Earth}}$). The gap between those groups \citep{fulton17}, known as the radius valley, is populated by a less frequent group of planets with ambiguous nature. These different types of exoplanets experience geological processes at extremes of pressure, temperature, and states of matter. Experimentally-grounded modeling frameworks are required to interpret the observed planetary properties \citep{Guimond2024,Lichtenberg2025}.

Atmospheres provide the principal means of detailed characterization of processes occurring on and within exoplanets \citep{Kempton2024}. The atmospheres of gas giants grow by the accretion of hydrogen and helium gas (H/He hereafter) from the circumstellar disk during the planet formation process, forming a primary envelope. In contrast, low-mass exoplanets that are similar to the terrestrial planets of the Solar System have a secondary atmosphere, thought to be produced by material exchanged with the planet’s deep interior through chemical reactions and transport of gases such as water (H$_2$O), carbon dioxide (CO$_2$), methane (CH$_4$), and sulfur dioxide (SO$_2$) into and out of the planetary mantle (ingassing and outgassing, respectively). These reactions couple the long-term climate and surface conditions of low-mass exoplanets to their interiors, establishing a connection that is distinct from that of gas giants \citep{Lichtenberg2025,Kempton2024}. 

It is unknown whether intermediate planets, such as sub-Neptunes and super-Earths, have primary envelopes or secondary atmospheres shaped by interior processes.  The JWST is providing detailed spectroscopic studies of the atmospheres and surfaces of low-mass exoplanets, enabling the application of geological models to exoplanet data. Comparing those observations to models provides information about the planets' bulk composition, internal dynamics, and overall geophysical history, bringing insights into their geology. This  requires constraining the surface and internal processes of exoplanets through atmospheric spectra, bulk density measurements, and host star compositions. 

\section*{Measuring exoplanet density}

The structure of a planet determines how its interior communicates with its surface and atmosphere. A planet’s structure and composition depends on its density, which constrains whether it is mainly composed of high-density iron, moderate-density rock, or lower-density ice or gas. Density is determined by measuring a planet’s mass [from radial velocities (RVs) or transit-timing variations] and radius (from transits). Short-period super-Earth masses have been measured to $\sim$15\% precision around bright low-mass stars (those classified as M dwarfs) and to $\sim$25\% around more massive Sun-like stars \citep{Pepe2021}. Determining the masses of even smaller planets with RVs is limited by stellar variability \citep{Hara2023}.  

Precise planetary radius measurements are provided by transits (typically) observed with space telescopes. The Kepler spacecraft, the Transiting Exoplanet Survey Satellite (TESS), and the CHaracterizing ExOPlanets Satellite (CHEOPS) have observed thousands of exoplanets and provided radii measurements with 10 to 15\% precision, although many of those systems are too distant for detailed follow-up. Determining planet radii from transit depth measurements requires knowing the host star radius, for which the Gaia spacecraft \citep{Vallenari2023} has provided stellar observations that enabled the precision of exoplanet host star radius determination to be improved by almost an order of magnitude \citep{MacDougall2023}.  

\section*{Distinguishing distinct exoplanet populations}

Precise exoplanet radius measurements demonstrated the bimodal radius distribution of sub-Neptunes and super-Earths discussed above. This bimodality was theoretically predicted \citep{Owen2013,Lopez2013,Sheng2014} to arise from planets with dense, rocky interiors and an initial hydrogen-dominated primary envelope, some of which later lost their atmospheres because of stellar irradiation, producing the smaller and denser planets we observe. Alternative explanations for the loss of the primary atmosphere include rapid boil-off immediately after formation \citep{Tang2024}, long-lived photoevaporation \citep{Owen2017}, or internal-luminosity-driven mass loss \citep{Ginzburg2018}. These models predict that super-Earths are mostly bare rocks or have secondary atmospheres and that sub-Neptunes have similar interiors but retain overlying thick hydrogen-dominated envelopes. Specifically, the interiors of both types of planet are predicted to be mixtures of rock and iron because any wider diversity in composition would smear out the resulting bimodal radius distribution, muting the distinction between populations. These models of the bimodal distribution do not require migration of the planets from farther out in the protoplanetary disk to their current short orbital periods, although migration is typically predicted with planet-formation models \citep{Drazkowska2023}.

\begin{figure*}[tbh]
\centering
\includegraphics[width=0.75\textwidth]{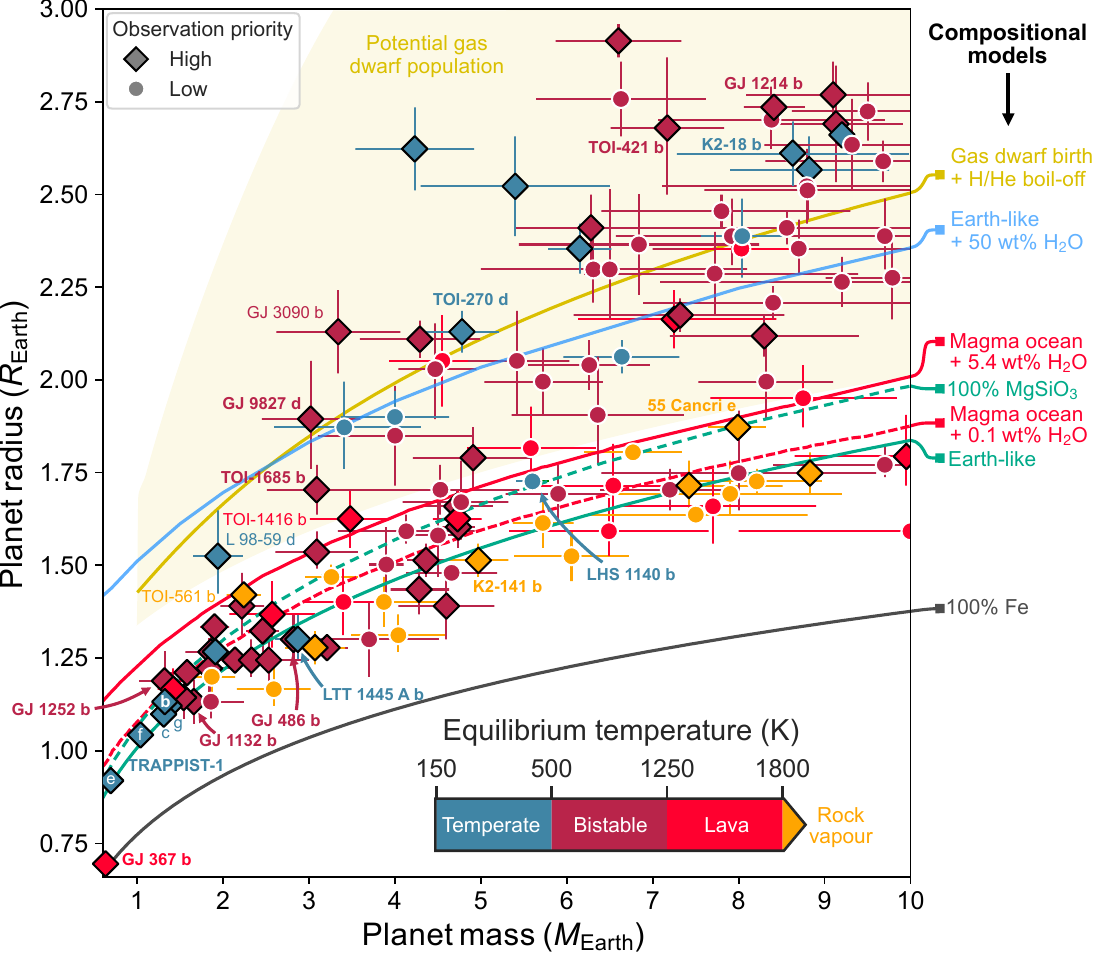}
\caption{\textbf{Observed exoplanet masses and radii compared with compositional models.} Radii [in Earth radii ($R_{\rm{Earth}}$)] are plotted as a function of mass [in Earth masses ($M_{\rm{Earth}}$)]. Colored symbols are observed exoplanets; we selected those with mass uncertainties $<$50\% and radius uncertainties $<$30\%, from \citet{Parc2024}. Symbol color indicates the equilibrium temperature, assuming full heat redistribution and zero albedo. Temperature thresholds in the color bar were chosen to reflect the regimes identified in  Fig. \ref{fig:regimes}. Curved lines are theoretical mass-radius relations for different compositions, as labeled \citep{Zeng2019,Dorn2021,Rogers2023}. Diamonds are planets with high transmission spectroscopy metric (TSM) or emission spectroscopy metric (ESM), making them a high priority for observational characterization; we selected those with TSM $>$ 90 if the planet radius $\geq 1.5$ $R_{\rm{Earth}}$, or TSM $>$ 10 for smaller radii, or ESM $>$ 7.5 \citep{2018PASP..130k4401K}. Some specific exoplanets are labeled; those in boldface are shown in Figs. \ref{fig:transmission_spectra} and \ref{fig:brightness_temperature}. The shaded region is a gas dwarf model of sub-Neptunes \citep{Rogers2023}, which assumes an Earth-like silicate mantle and iron core composition, surrounded by a H/He envelope accreted from the protoplanetary disk.
}
\label{fig:M-R}
\end{figure*}

The radii and masses of the precisely characterized exoplanets are shown in Fig.~\ref{fig:M-R}, selected to have $<$ 3~$R_{\rm{Earth}}$ and $<$ 10 Earth masses. We used the planetary equilibrium temperature, defined as the theoretical temperature of blackbody emission in radiative equilibrium with the global heat redistribution (which is always lower than the substellar temperature of a tidally locked blackbody), to define four broad climate regimes within this parameter space \citep{Lichtenberg2025}. The onset of a runaway greenhouse effect, when ocean water is expected to evaporate in a feedback loop at increasing stellar irradiation \citep{Hamano2013,Kopparapu2013}, separates the regimes we label as temperate and bistable in Fig.~\ref{fig:M-R}. Models predict that any planets in the bistable regime that have substantial water (in their atmosphere or on their surface) do not have a stable temperate climate, but instead experience a super-runaway climate. In the super-runaway regime, radiative equilibrium is reached only after all oceans are evaporated and persists until atmospheric escape removes most of the atmosphere. However, variations in starting configuration (for example, the presence of an underlying magma or water ocean) affect the climate model predictions \citep{Turbet2021,Boer2025}. At atmospheric pressures similar to the Solar System terrestrial planets, solid rock begins to melt at $\approx$1250 K, which we label as the lava planet regime. At even higher temperatures ($\gtrsim$1800 K), molten surfaces become increasingly vaporized, forming a thin veneer of gaseous magnesium (Mg), sodium (Na), and silicon (Si) around the planet, which we label as the rock vapor regime.

The measured mass and radius of each exoplanet can be reproduced by an entire family of structural models \citep{seager07,valencia07}, so each exoplanet has multiple degenerate explanations with varying compositions and internal structures \citep{rogers10,Dorn2015}. For example, the lava world 55 Cancri e can be explained equally well by models of either 100\% MgSiO$_3$ rock or by an Earth-like internal structure with a molten rock mantle (magma ocean) and $\approx$5 wt\% bulk water content (the latter is more than two orders of magnitudes greater than that of Earth) \citep{Dorn2021}. 

We also show in Fig. \ref{fig:M-R} a population of intermediate-density planets around $\sim$2~$R_{\rm{Earth}}$ that span a wide range in temperature. These planets are rare around Sun-like stars \citep{fulton17}, but have the potential to constrain models of the formation and evolution of small planets. Planet formation theory predicts that planets typically start accreting outside of the water ice line (the boundary of the region where solid water ice can survive in the protoplanetary disk) \citep{Drazkowska2023}, followed by migration into closer orbits \citep{Zeng2019,Lichtenberg2019}. This process is predicted to produce at least some short-period planets that incorporated volatile ice-rich solids \citep{Venturini2020,Izidoro2022}. It is therefore possible that a fraction of the larger 1 to 3~$R_{\rm{Earth}}$ planets are highly water-rich, because they originated from the outer regions of their protoplanetary disk \citep{Luque2022,Burn2024}. However, high-molecular-weight volatiles (C-N-S-rich gas and ice compounds) in planet atmospheres could alternatively arise from preferential loss of hydrogen from the atmosphere \citep{Chen2016,Malsky2023} or from chemical reactions between molten metal or silicate on the surface and H/He-rich gaseous envelopes \citep{Kite2021,Lichtenberg2021,Schlichting2022}. 

These examples illustrate that mass and radius alone are not enough to distinguish between different structures and bulk compositions. Additional information is required, which can be provided by measurements of the atmospheric composition. 

\section*{Transmission spectroscopy of exoplanet atmospheres}

\begin{figure*}[tbh!]
\centering
\includegraphics[width=0.75\textwidth]{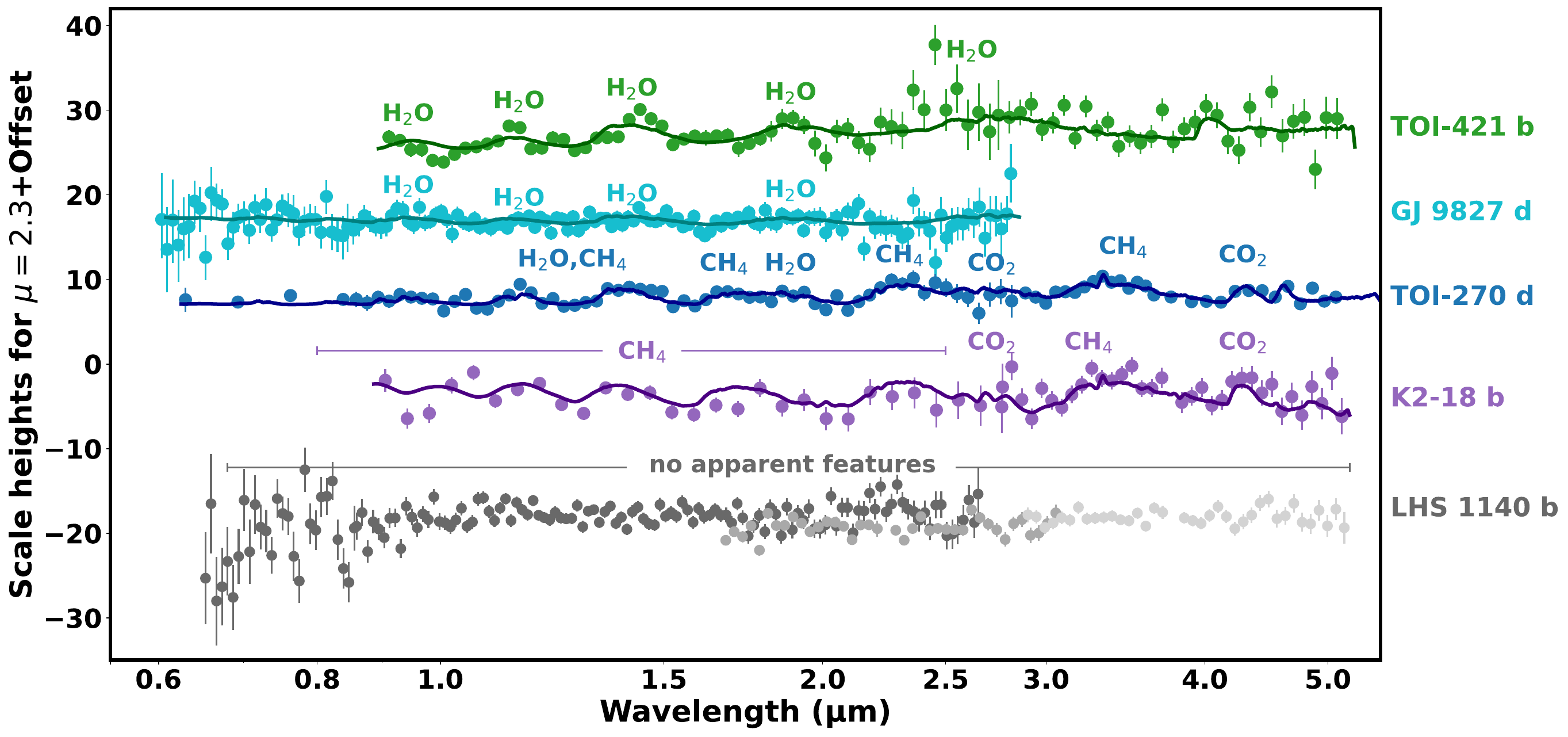}
\caption{\textbf{Transmission spectra and atmospheric models for selected small exoplanets observed with JWST.} Data points are the observed spectra of five example exoplanets discussed in the text, with vertical bars indicating the 1-$\sigma$ uncertainties. Thick solid lines indicate the best-fitting atmospheric model, as reported by each study. Colors indicate different exoplanets: green is TOI-421 b \citep{Davenport2025}, cyan is GJ 9827 d \citep{Piaulet-Ghorayeb2024}, blue is TOI-270 d \citep{Benneke2024}, purple is K2-18 b \citep{Madhusudhan2023}, and gray is LHS 1140 b (JWST/NIRISS in dark gray and JWST/NIRSpec in light gray) \citep{Cadieux2024,Damiano2024}. The planets are ordered from hottest ($\sim$920~K) to coldest ($\sim$~225~K) equilibrium temperature, assuming zero albedo; they also have different masses and radii (Fig.~\ref{fig:M-R}). For comparison, the spectra are offset by a constant value and have been converted to atmospheric scale height, assuming a mean molecular weight ($\mu$) of $2.3$ g/mol. This weight is unlikely to be valid for all these exoplanets, given their diversity of compositions. Labels and thin horizontal lines above each spectrum identify the molecules responsible for absorption features (which extend upward in this representation). The wavelength scale is logarithmic.
}
\label{fig:transmission_spectra}
\end{figure*}

Observations made with JWST have provided atmospheric composition measurements for a diverse range of exoplanets. Some of the molecular features observed in JWST transmission spectra of small ($\approx$1.7 to 2.8~$R_{\rm{Earth}}$) exoplanets are shown in Fig.~\ref{fig:transmission_spectra}. These transmission spectra record starlight that has been filtered through the exoplanet's atmosphere and are plotted in terms of scale height, the characteristic length scale of an atmosphere. The labeled features arise from absorption by H$_2$O, CH$_4$, and CO$_2$. 

The strengths of these features constrain the relative abundances of the detected molecules, which Fig.~\ref{fig:transmission_spectra} shows to be temperature dependent. This trend was identified using Hubble Space Telescope (HST) observations \citep{Brande2024}, which showed weak spectral features in the 500 to 700~K equilibrium temperature range. This was interpreted as being due to cloud condensation or photochemical hazes produced by photolysis of CH$_4$, which is expected to be the dominant C-bearing species at those temperatures. At hotter temperatures, carbon monoxide (CO) is expected to be the dominant C-bearing species, so hazes are predicted to be less common. At cooler temperatures, the observations show little to no obscuration by aerosols (clouds or hazes) -- for example the spectra of K2-18~b \citep{Madhusudhan2023,Luque2025b} and TOI-270~d \citep{Benneke2024,Holmberg2024} in Fig.~\ref{fig:transmission_spectra} are consistent with cloud-free atmospheres. 

If small planets have atmospheres dominated by H/He, planets in the 500 to 700~K range are expected to show featureless transmission spectra due to efficient cloud and haze formation. However, the intermediate-temperature planet GJ 9827~d provides a counter-example: its spectrum is not featureless and provides no evidence of clouds. Instead, its spectrum is consistent with an H$_2$O-dominated atmosphere, with a volume mixing ratio $>$31\%, and an O/H ratio of $\approx$4 by mass. At this planet's equilibrium temperature, water is not expected to condense in the atmosphere or at the surface, and hydrogen should be well mixed with water, so the upper-atmosphere O/H is representative of deeper layers \citep{Piaulet-Ghorayeb2024}. This atmospheric observation breaks the degeneracy between models of low-metallicity (low abundance of elements heavier than He) cloudy atmospheres and high mean-molecular weight atmospheres (expected to produce weak spectral features due to small scale heights), indicating that GJ 9827~d is $>$20\% water by mass. This could arise if the planet formed from material with a high volatile ice/rock ratio \citep{Zeng2019,Lichtenberg2019,Venturini2020}. Alternatively, it is possible to produce a water-rich atmosphere through geochemical interactions between a (perhaps primordial) hydrogen-rich atmosphere and the planet's magma ocean during planetary accretion \citep{Ikoma2006,Kite2021,Kimura2022,Rogers2024}. Determining the origin of volatiles in small planet atmospheres require larger samples of exoplanets spanning the density-irradiation parameter space \citep{Schlecker2024}.

Two other planets in Fig.~\ref{fig:transmission_spectra} provide evidence of mixing and interaction with some kind of surface underneath the atmosphere. The JWST spectrum of K2-18~b indicates the presence of CH$_4$ but not NH$_3$ \citep{Madhusudhan2023}, although pure-atmosphere models expected both in a sub-Neptune planet of K2-18~b's temperature ($\sim$250~K). The low NH$_3$ abundance could either be explained by chemical dissolution of N-bearing species into an internal magma \citep{Shorttle2024,Glein2025} or water ocean \citep{Yu2021,Hu2021,Tsai2021}. The statistical significance of CO$_2$ and other atmospheric compounds in K2-18 b's atmosphere are under debate \citep{Luque2025b,Welbanks2025}. The spectrum of TOI-270~d shows features of CH$_4$ and CO$_2$, which can be explained by a high atmospheric mean-molecular weight. This finding is inconsistent with a distinct H/He layer above a metal/rock-rich mantle \citep{Benneke2024}. Instead, TOI-270~d's envelope could contain similar amounts of C- and O-rich volatiles and H/He mixed together in a supercritical phase. As with GJ 9827~d above, this scenario can arise either from magma ocean-atmosphere interactions or initial enrichment of volatile ices. These examples illustrate how JWST observations constrain interior-atmosphere interactions. 

About a dozen more small planet JWST transmission spectra are featureless, which could be due to atmospheres that are either (a) of high enough mean-molecular weight that features of individual species are below the JWST detection limit, (b) blanketed by high altitude clouds or hazes that mute molecular features, or (c) no atmospheres at all, just a solid or liquid (magma) surface. Among these, scenario (b) is a plausible explanation for most featureless transmission spectra. The higher-precision spectrum of the sub-Neptune GJ 1214~b is especially constraining; it indicates that both high mean-molecular weight and high-altitude aerosols act together at least for this planet \citep{Gao2023}. Scenario (c) is possible only for smaller planets with bulk densities that are consistent with no atmospheric envelope. 

Featureless spectra also constrain the pressure of the cloud (or haze) top at a given atmospheric composition or metallicity. Atmospheric metallicities of $\gtrsim$200$\times$ that of the Sun are required to explain the lack of observed spectral features \citep{Gao2023}. Most of the small planets observed with JWST transmission spectra orbit M-dwarf stars, which have cool photospheres, which is a star's outer shell from which light is radiated, and a high prevalence of stellar spots, which are darker and cooler regions of strong magnetic activity that suppress convection. These factors can contaminate the planet transmission spectrum because an M-dwarf's photosphere can also host water, and the occulted part of the stellar disk could differ from the average stellar spectrum. Several of the JWST small planet transmission spectra indicate such contamination, particularly the Earth-sized planets orbiting the star TRAPPIST-1 \citep{Trappist2024}. Therefore, transmission spectra are insufficient to determine whether these planets have atmospheres and what they are made of.

\section*{Eclipse spectroscopy of atmospheres and surfaces}

Secondary eclipse spectroscopy, which compares observations immediately before and after the planet moving directly behind the star, has fewer issues with host star contamination. These observations determine the dayside emission from the planet (the side that is facing the star), which is permanently fixed because transiting planets are expected to be tidally locked to their host star \citep{Barnes2017,Farhat2025}. If the atmosphere is sufficiently dense, heat from the irradiated dayside can be circulated to the night-side, causing the dayside flux measured from eclipse observations to be lower than expected for a blackbody at the equilibrium temperature of the planet \citep{Mansfield2019,Koll2019}. Another method, which requires more telescope time, is to observe emission from a planet over its full orbit -- a phase curve -- which can constrain night-side emission due to atmospheric circulation \citep{Koll2019,Hammond2025}. If there is no atmosphere, the dayside emission can provide information about the planet's surface composition, mineralogy, and geology \citep{Guimond2024}. 

\begin{figure*}[tbh!]
\centering
\includegraphics[width=0.73\textwidth]{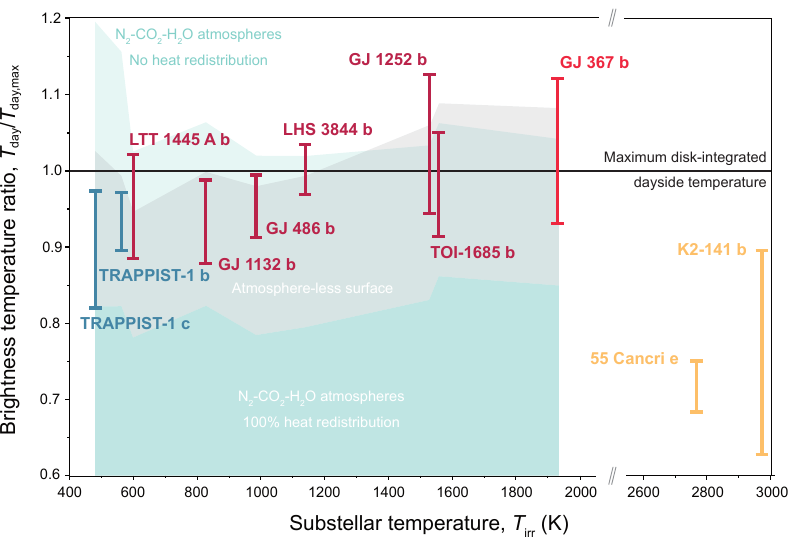}
\caption{\textbf{Brightness temperature ratio from eclipse observations compared with theoretical models.} Colored bars are the 1-$\sigma$ ranges of brightness temperatures ($T_\mathrm{day}$) derived from eclipse observations, normalized to the theoretical maximum disk-integrated dayside temperature for a zero-albedo, zero-heat redistribution planet (solid black line). They are plotted as a function of the substellar temperature of a tidally locked planet with zero albedo at all wavelengths ($T_\mathrm{irr}$). Bar colors indicate the equilibrium temperature categories defined in Fig.~\ref{fig:M-R}, calculated by using published methods \citep{ParkCoy2024} applied to eclipse depths and uncertainties from JWST and Spitzer emission observations as reported in the literature. Data sources are TRAPPIST-1 c \citep{2021PSJ.....2....1A,2023Natur.620..746Z}, TRAPPIST-1 b \citep{2021PSJ.....2....1A,2023Natur.618...39G,2024NatAs.tmp..292D}, LTT 1445 A b \citep{Wachiraphan2025}, LHS 3844 b \citep{Kreidberg2019,2019ApJ...871L..24V}, GJ 1132 b \citep{2024ApJ...973L...8X}, GJ 486 b \citep{2024ApJ...975L..22W}, GJ 1252 b \citep{2020ApJ...890L...7S,2022ApJ...937L..17C}, TOI-1685 b \citep{Luque2025,2023ApJ...955L...3G}, GJ 367 b \citep{2024ApJ...961L..44Z}, 55 Cancri e \citep{2024Natur.630..609H,Patel2024}, and K2-141 b \citep{2022A&A...664A..79Z}. Shaded regions indicate the ranges predicted by theoretical models for the specific exoplanets with red and blue bars \citep{Hammond2025}; gray is a model of bare rock surfaces with no atmosphere. Cyan is a model of atmospheres with N$_2$-CO$_2$-H$_2$O compositions with 1 to 10 bar total surface pressure, with (top) no (light cyan) or (bottom) complete (dark cyan) day-to-night heat redistribution through atmospheric circulation.}
\label{fig:brightness_temperature}
\end{figure*}

Distinguishing planets with bare surfaces from those with atmospheres is not straightforward. Fig. \ref{fig:brightness_temperature} shows the ratio of observed dayside brightness temperature ($T_\mathrm{day}$), a measure of blackbody temperature to duplicate the observed intensity, to the theoretical maximum disk-integrated dayside temperature for a zero-albedo, zero-heat redistribution planet ($T_\mathrm{day,max}$), of 11 transiting rocky exoplanets measured from JWST secondary eclipse observations, as a function of the substellar temperature of a tidally-locked planet with zero albedo at all wavelengths ($T_\mathrm{irr}$). Most of the secondary eclipse observations (those with $T_\mathrm{irr} \lesssim $ 2000~K) are consistent with either low-albedo bare-rock surfaces (i.e. no atmosphere), or alternatively high mean-molecular weight atmospheres with surface pressures $\lesssim$10 bar and with little heat redistribution by atmospheric circulation. The degeneracy is exacerbated by uncertainty in stellar spectral models \citep{Tayar2022}, which are used to calculate the brightness temperatures \citep{ParkCoy2024,Luque2025}, and possible interactions of secondary atmospheres with the interiors of planets, which could store a substantial amount of volatiles \citep{Dorn2021}, and thus reduce the outgassed atmospheric envelope. There is no evidence for strong heat redistribution or thick cloud-free atmospheres in this observed sample, but nor is it clear whether these rocky exoplanets orbiting M-dwarf stars have secondary atmospheres of $\lesssim$10 bar surface pressure. Intense stellar X-ray and EUV radiation of M-dwarf stars gradually strips the gaseous atmosphere from orbiting planets, but the efficiency of this process is highly sensitive to the composition of the atmosphere \citep{Chatterjee2024}. 

\section*{Using atmospheres to study planetary interiors}
\label{sec:interior}

Small and highly irradiated exoplanets are most likely to have had their atmospheres removed, allowing observations to directly probe their surfaces. Otherwise, mass exchange between the interior and surface of a planet is expected to be recorded in the composition of the atmosphere. The transport of mass and energy between the internal and external layers of planets is driven by the redistribution of heat provided by accretion, decay of radioactive elements, tidal interactions, and stellar irradiation; the latter two sources are particularly relevant for short-period planets.

Exchange of material between interiors and atmospheres imprints information about the interior of a planet in the chemistry of its atmosphere. The interiors are at higher pressures and temperatures than the atmospheres. In the relative cold and low pressure of an atmosphere, the chemical reactions that occur in the interior are likely to become quenched, leaving long-lasting and detectable signs of disequilibrium chemistry; such transport from lower atmospheric layers has been proposed to explain the presence of NH$_3$ in the atmospheres of sub-Neptunes \citep{Yang2024}. This atmospheric disequilibrium could drive reactions between the transported material and the background atmospheres, leading to the sequestration of gases from an atmosphere into the deeper layers of the planet. This is analogous to atmospheric CO$_2$ and O$_2$ on Earth, which combine with liquid water and rock, producing carbonates and oxides, respectively, in Earth's crust and mantle \citep{Sleep2005}. This forms part of a geological cycle, potentially returning to the atmosphere hundreds of Myr later \citep{Dasgupta2010}. Indications of interior-atmosphere exchange could be enhanced if mass transport occurs across a compositional stratification inside a planet; for example on Earth volcanic eruptions release $\mathrm{SO}_2$ into the atmosphere, which react with water vapor to form aerosols. The presence or absence of an atmospheric gas species can therefore provide insight into a planet's structure and dynamics.

\begin{figure*}
\centering
\includegraphics[width=0.87\textwidth]{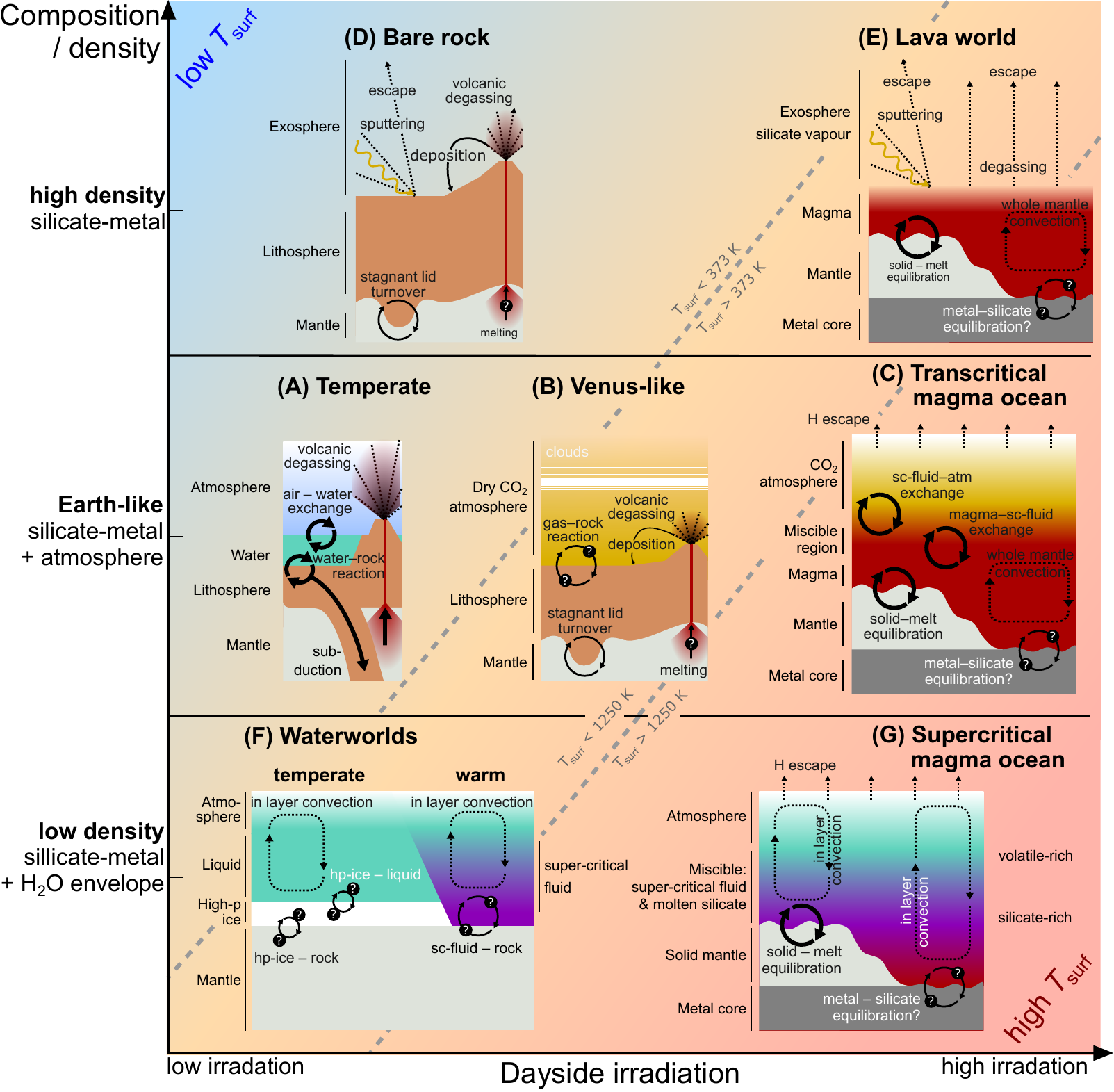}
\caption{\textbf{Schematic illustration of the geodynamic regimes predicted for planets with different densities and dayside irradiations.} (\textbf{A} to \textbf{G}) Seven regimes discussed in the text are illustrated at qualitative locations of density as a function of dayside irradiation by the host star. Each regime is represented by a schematic illustration of the processes that exchange mass and energy between planetary reservoirs; arrows indicate small (thin) or large (thick) fluxes; question marks indicate poorly constrained processes (for example, interactions between rock and high-pressure ice). Background color shading qualitatively indicates the expected temperature at the atmosphere-interior interface. In (A) to (G), the left axis indicates the structural interior layers of the planet, and colors indicate composition. The diagonal gray dashed lines indicate where the predicted surface temperature ($T_{\rm{surf}}$; at the interface where the atmosphere or envelope contacts silicate material) equals those of boiling water and silicate melting, labeled with their values at 1 bar atmospheric pressure. Thicker atmospheres produce higher surface temperatures for a given irradiation, shifting the irradiation at which a change in geodynamic regime occurs.}
\label{fig:regimes}
\end{figure*}

The nature of interior-atmosphere exchange on a given planet depends on its geodynamic regime, the mode of mass and energy transport it experiences. Earth's mobile-lid geodynamic regime results in plate tectonics on its surface, but the origin of that regime is debated \citep{Brown2020}, and it is unclear which exoplanets experience similar geodynamics \citep{Meier2021,Meier2024}. Two factors probably set the long-term geodynamic processes on exoplanets: their density, which is a proxy for their composition; and their stellar irradiation (instellation), which determines their thermodynamics through the ratio of in- and outgoing energy fluxes. The combination of irradiation and composition, for a given mass of an exoplanet, provide a broad description of a planet's structure and dynamics, because they determine the interfaces between interior layers and how material is transported between them. Fig. \ref{fig:regimes} presents a schematic view of predicted exoplanetary thermodynamic climate and interior regimes derived from laboratory measurements, observations, and empirical evidence in geochemistry, geophysics, climate science, planetary science and astronomy \citep{Guimond2024,Kempton2024,Lichtenberg2025}. Those interfaces can be abrupt (e.g., the rocky surfaces of terrestrial planets) or diffuse (e.g., the edges of atmospheres).  The pressure, temperature, and composition of matter on either side of each interface determine the rate of exchange of mass and energy between the layers. The predicted diversity of geodynamic regimes for planets across the density--instellation parameter space (Fig. \ref{fig:regimes}) indicates how exoplanet interior structures can potentially be constrained by observing their atmospheres.

\section*{Super-Earths and terrestrial exoplanets} 

Planets close to the Earth-like composition line in Fig. \ref{fig:M-R} are predicted to exhibit geodynamic processes which are familiar to planetary scientists, so we consider them first.  

In the temperate regime (Fig. \ref{fig:regimes} A), aqueous chemistry mediates the long-term atmospheric expression of interior-atmosphere exchange and regulates heat transport. Exchange between the atmosphere and the liquid-water hydrosphere is expected to be rapid, by analogy to Earth's oceans, which ventilate in $\lesssim$5000 years and rapidly equilibrate with the atmosphere. On these planets, gases that readily dissolve in liquid water, such as $\mathrm{SO}_2$, are removed from the atmosphere within a few years \citep{Loftus2019}, so are only transiently present following major inputs, such as large volcanic eruptions.  Atmospheric $\mathrm{CO}_2$ is depleted by water-rock reactions, which can occur at sharp interfaces on exposed landmasses and at the base of the oceans. These reactions drive Earth's climate thermostat; we therefore expect that $\mathrm{CO}_2$ would be present at low concentrations in the atmospheres of habitable terrestrial worlds \citep{Walker1981,Foley2015, Krissansen-Totton2018}. Planets in this regime would have low atmospheric abundances of some volatile elements ($\mathrm{CO}_2$, $\mathrm{SO}_2$, $\mathrm{H}_2\mathrm{O}$) that are otherwise cosmochemically abundant.

At high irradiation, terrestrial planets are expected to experience a climate collapse (Fig. \ref{fig:regimes} B, Venus-like). Evaporation of liquid water oceans would produce a steam-rich atmosphere; the atmospheric water is then photodissociated and hydrogen lost to space \citep{Luger2015,Schaefer2016}. This runaway greenhouse scenario \citep{Simpson1928,Nakajima1992} appears to have occurred on Venus \citep{Constantinou2024}, either early \citep{Hamano2013} or late \citep{Way2016} in its history. The result is a planet with a dry and oxidized atmosphere with high mean-molecular weight, dominated by the atmospheric chemistry of carbon- and sulfur-bearing species \citep{Jordan2025}.  On these planets, the atmosphere-to-surface interaction is mediated by reactions between gas and rock, which are less efficient at transferring material than aqueous reactions \citep{Byrne2024}. In this regime, volcanic input to the atmosphere is potentially suppressed by two factors: First, a massive $\mathrm{CO}_2$ atmosphere provides a high surface pressure that would resist degassing because gases are more soluble in magmas at higher pressure \citep{Suer2023}. Second, the geodynamic feedbacks of a hot, dry planetary surface is expected to lower overall volcanic activity by inhibiting plate tectonics \citep{Nimmo1998}. We expect planets in this regime to have wholly volcanically-derived atmospheres, but of an ancient nature, and a clearly demarcated rock--atmosphere interface.

All rocky planets likely pass through a magma ocean period as they dissipate the initial heat from their formation \citep{Lichtenberg2023,Lichtenberg2025}. However, highly irradiated rocky planets could be trapped in this state permanently (Fig. \ref{fig:M-R} C, Transcritical Magma Ocean). These planets would retain a high mean-molecular weight atmosphere, which is H-poor due to the same processes as in Venus-like climates, directly overlaying a molten mantle.  Gas exchange with magma is far more efficient than with rock, and rapid convection within the magma itself would facilitate exchange between the atmosphere and interior \citep{Kite2016,Lichtenberg2021JGRP,Bower2022}. Because the interior is vastly more massive than the atmosphere in this regime, the atmospheric chemistry would reflect the interior chemistry \citep{Maurice2024,Nicholls2025}; for example, redox chemistry and the relative abundances of gas species would be dictated by their solubility in the magma \citep{Suer2023,Boer2025}. Depending on the instellation and atmospheric opacity, the magma ocean could extend down to the core-mantle boundary. Sufficiently massive atmospheres would become supercritical fluids at their base, leaving no clear boundary with the magma \citep{Pierrehumbert2023,Young2024}. In this regime, chemical exchange could in principle extend through the entire planet -- including keeping the metal core in chemical equilibrium with the mantle and atmosphere \citep{Lichtenberg2021,Schlichting2022}. We expect planets in this regime to have atmospheres that directly reflect their interiors \citep{Lichtenberg2025}.

\section*{Atmosphere-stripped worlds}

Planets with bulk densities approximately similar to or greater than Earth (below the 100\% MgSiO$_3$ or Earth-like lines in Fig. \ref{fig:M-R}) might have lost their initially accreted volatiles to space. The surfaces of such planets are directly subject to intense stellar irradiation, which can drive their geodynamic and petrologic evolution into regimes unknown in the Solar System. Planets that completely lose their atmospheric volatile inventory are expected to be driven by tectonic and geological processes that span their lithosphere (Fig. \ref{fig:regimes} D, Bare Rock). Energy transport across this layer is predicted to be dominated by the rheological properties of the mantle, internal energy sources, and irradiation. Bare-rock exoplanets that orbit close to their host star tend to become tidally locked \citep{Farhat2025}, producing alternative forms of geodynamic evolution that are driven by the strong day-night temperature contrast \citep{Meier2021,Meier2024}. The surface composition of these exoplanets could reflect their petrological past \citep{Hu2012,Kreidberg2019}.

Atmosphere-less planets that are subject to intense irradiation \citep{Kite2016} or strong tidal forces \citep{Farhat2025} from their host star are expected to have molten surfaces and interiors. These could potentially be observed as a global magma ocean or dayside magma pool (Fig. \ref{fig:regimes} D, Lava Worlds). Energy and mass exchange in these systems is predicted to be highly dependent on the geometry and heat transport of the magma ocean or pool, dynamical interaction with the underlying iron core, and transition between magma and rock surface (magma ocean shorelines) across the terminator \citep{Boukare2022,Meier2023}. On these planets, the outgassing of refractory compounds can indicate the underlying magma composition, and is potentially observable \citep{Piette2023,Zilinskas2023,Wolf2023}.

\section*{Sub-Neptunes: water- or magmaworlds}
The low bulk densities of sub-Neptune exoplanets imply the presence of substantial volatile envelopes (Fig. \ref{fig:regimes}). The envelope could be composed of primary accreted H/He gas \citep{Owen2017,2022ApJ...941..186L}, or be $\mathrm{H}_2\mathrm{O}$-dominated due to the planet having formed outside the water ice line \citep{Venturini2020,Burn2024}. The common feature of their interiors is that chemical exchange between silicates and volatiles happens at high pressure ($\gtrsim$ GPa). This leads to different forms of mass and energy exchange than on terrestrial planets, producing different global dynamics and thermal histories \citep{Vazan2018,Kite2020}.

For moderate pressures at the envelope-rock interface ($<$10\,GPa) and low temperatures ($\approx$1000\,K, at low instellation) in a deep envelope of $\mathrm{H}_2\mathrm{O}$, it is predicted that rock is in contact with high-pressure water ice (Fig. \ref{fig:regimes} F, Temperate Waterworlds). Energy transport out of the rocky interior would therefore be modulated by convection in the ice layer, at a rate that depends on the ice viscosity. The rheological properties of the ice are affected by its composition and the incorporation of other materials into the $\mathrm{H}_2\mathrm{O}$-ice structure. $\mathrm{CO}_2$ transport across the high-pressure ice layer is predicted to be limited \citep{Levi2023}; the solubility of silicate components in high-pressure ice is largely unknown. The chemical and geodynamic feedbacks between ice composition and the properties of the overlying ocean and atmosphere are also poorly understood \citep{Journaux2020}.

At even moderate instellation, due to the insulating effect of a thick volatile envelope (whether composed of $\mathrm{H}_2\mathrm{O}$ or H/He), high-pressure water and hydrogen layers are in a supercritical fluid state (Fig. \ref{fig:regimes} F, Warm Waterworlds). If the temperature is below the melting point of rock, this produces a sharp boundary beneath supercritical $\mathrm{H}_2$ or $\mathrm{H}_2\mathrm{O}$ \citep{Pierrehumbert2023,Young2024}. The chemistry of reactions between supercritical fluids and rock is not well understood, including the kinetics and capacity of material transport \citep{Innes2023}. Because transitions in supercritical fluids are continuous, this regime could involve compositional stratification, with deeper fluids enriched in silicate components underlying more water-rich layers above \citep{ElkinsTanton2008}. This is expected to reduce the transport of mass and energy between the interior and atmosphere \citep{Leconte2024}.

Higher levels of instellation produce molten rock at the base of the ice or gaseous envelope. Such a planet is predicted to not have any clearly defined internal layers (Fig. \ref{fig:regimes} G, Supercritical Magma Ocean), because molten silicate is miscible in high-pressure-fluids \citep{Vazan2022,Kovacevic2022}. This allows rapid reactions between the constituents, such that thermochemical equilibrium prevails throughout the continuous fluid envelope of the planet. Depending on the vertical extent of the supercritical regime \citep{Young2024}, strong gradients in mean-molecular weight of the fluid could occur, with a silicate-rich wet magma at depth, and water-rich material at lower pressures. Sub-Neptunes in this regime are expected to have well-mixed, supercritical deep atmospheres, rather than stratified H$_2$O layers underlying H$_2$ \citep{Pierrehumbert2023,Nixon2024}. These compositional transitions affect heat and mass transport through the planet, but in this regime the silicate interior becomes chemically coupled to the atmosphere, even if the silicate components rain out at much higher pressure levels \citep{Vazan2023}. This is predicted to have consequences for the observed mass-radius relation \citep{Vazan2022} and upper-atmosphere abundances \citep{Misener2023,Ito2025}.

\section*{Interior-atmosphere exchange on hot exoplanets}

Transit observations are biased towards detecting close-in, warm to hot exoplanets (Fig. \ref{fig:regimes}, B, C, E \& G). These span transitions between planetary states that are predicted to show global chemical and energy exchange, a regime that is not present in the present-day inner Solar System, but could reflect the Solar System planets in their early evolution following accretion \citep{Lichtenberg2023}. JWST spectroscopy (Fig.~\ref{fig:transmission_spectra}) can be interpreted as showing that the sub-Neptunes K2-18 b and TOI-270 d are either cold, water ocean-bearing sub-Neptunes \citep{Madhusudhan2023,Holmberg2024,Luu2024}, or have a supercritical lower envelope and possible magma ocean surface \citep{Shorttle2024,Wogan2024,Glein2025}. An analogous dichotomy of interpretations applies to Venus, which might have had a water ocean in its early evolution \citep{Way2016} or instead lost its water during a magma ocean and super-runaway greenhouse state that lasted hundreds of Myr \citep{Turbet2021,Constantinou2024}. Distinguishing between these options, for both Venus and sub-Neptune exoplanets, is necessary to determine whether the surfaces were ever habitable \citep{Lichtenberg2025}. 

It remains uncertain how much of a planet's inventory of volatile elements can be stored in the interior during extended magma ocean phases \citep{Dorn2021,Luo2024,Boer2025}; if interiors efficiently remove volatiles from atmospheres, then volatiles could remain present on the surface at low levels for Gyrs, rather than being rapidly lost to space \citep{Kite2020,Dorn2021,KrissansenTotton2024}. Efficient interior-atmosphere exchange would also strongly affect the atmospheric composition \citep{Nicholls2024,Boer2025}, so itself influences the atmospheric loss rate \citep{Chatterjee2024,Cherubim2025}. Nitrogen- and sulfur-bearing species are predicted to trace interaction with molten silicate mantles \citep{Nicholls2025,Jordan2025}, but are also affected by liquid water oceans \citep{Loftus2019,Hu2021}.

Evolutionary considerations could provide a way to break this degeneracy for K2-18 b, TOI-270 d, GJ 9827 d and similar exoplanets. The transition from a hot super-runaway regime to colder, potentially temperate climates is substantially different from the opposite cold to hot climate transition because of cloud feedbacks and volatile dissolution in magma oceans \citep{Turbet2021,Nicholls2025b}, so the initial conditions affect the evolution of each planet \citep{Lichtenberg2021JGRP,KrissansenTotton2024}. This climate transition is not symmetric \citep{Boer2025}, as had previously been assumed \citep{Kopparapu2013}. Climate models of H$_2$- and H$_2$O-dominated atmospheres \citep{Innes2023,Selsis2023,Leconte2024} indicate that super-runaway atmospheres have radiative, non-convective layers in their interiors, which reduce energy transport across the atmosphere-interior interface. The evolutionary effect of this on low-mass exoplanets depends on their volatile composition and geochemical interactions with the underlying magma ocean \citep{Nicholls2025,Nicholls2025b}. JWST observations of sub-Neptune exoplanets indicate that their atmospheres are composed of high mean-molar weight components, which are predicted to interact with the planetary interior (see above). More precise transit surveys, such as PLATO (PLAnetary Transits and Oscillations of stars) \citep{Rauer2025}, are required to empirically probe the radius--density transition that is predicted across the inner edge of the classical habitable zone \citep{Turbet2019,Schlecker2024}, and statistically test interior-atmosphere feedbacks on Earth-sized- and super-Earth exoplanets.

Additional spectroscopic observations are required to determine the C-N-S abundances in exoplanet atmospheres, which are affected by chemical and transport processes at the interface with the interior \citep{Liggins2023,Brachmann2025}. The speciation and solubility of C-N-S compounds are poorly constrained at chemically reducing (H-rich) and high pressure conditions \citep{Suer2023}. Experimental investigation of photochemical haze production and cloud particle properties in these regimes are necessary inputs for interpretative models \citep{Horst2018,Gao2021}. Progress will require interdisciplinary collaboration between astronomers, planetary scientists and geoscientists, combining observational and experimental data with models of the energetic and chemical exchange across the interior-atmosphere interface. 

\section*{Atmospheres on rocky exoplanets orbiting M-stars}

The experimental and theoretical evidence that atmospheres rapidly interact with magmatic and metallic liquids \citep{Suer2023,Lichtenberg2025} implies that planetary interiors have substantial capacity for storing volatile elements, particularly in the magma ocean and transition regimes (Fig. \ref{fig:regimes}, C \& G). It is unknown whether this also applies to Earth-sized rocky exoplanets. All observed transmission spectra of rocky exoplanets orbiting M-dwarf stars have been featureless (such as LHS 1140 b in Fig. \ref{fig:transmission_spectra}), indicating that they do not have large gaseous envelopes. Eclipse measurements of those bodies are generally consistent with a blackbody at the expected equilibrium temperature; this is compatible with either bare rock or thin atmospheres ($\lesssim$ 10 bar) (Fig.~\ref{fig:brightness_temperature}). That demonstrates that volatile escape to space is more rapid than volatile delivery to the gaseous envelope. Investigating this process as a function of orbital distance, stellar type, and planetary mass could potentially provide information on the prevalence of exoplanets with Earth-like atmospheres.

Connecting these results to planetary formation theory \citep{Drazkowska2023,Lichtenberg2023} will require spectroscopic observations of the emitting layer (whether it is an atmosphere or rocky surface), because the presence or absence of gaseous envelopes only indirectly reflects the interior volatile storage \citep{Guimond2024,Lichtenberg2025}. It is debated whether statistical samples of transmission spectra or secondary eclipse spectra provide better prospects for determining the prevalence of substantial atmospheres on these exoplanets \citep{Koll2019,Trappist2024,ParkCoy2024}, or if measuring the full phase curves of a smaller number of exoplanets would provide greater insight \citep{Hammond2025}.

\section*{Prospects for Earth-like exoplanets}

JWST can potentially constrain the link between the atmosphere and interiors of sub-Neptunes and super-Earths, and the prevalence of volatile envelopes on rocky M-dwarf exoplanets. However, it is unlikely to provide similar understanding of Earth-like exoplanets. Upcoming exoplanet surveys will probe different parameter spaces to JWST's. PLATO \citep{Rauer2025} will provide long-baseline transit monitoring campaigns, but with only highly limited capabilities for atmospheric spectroscopy. The Nancy Grace Roman Telescope will use gravitational microlensing to search for exoplanets of one Earth mass and below \citep{Tamburo2023}, which is expected to constrain the mass function of exoplanets, but microlensing does not determine planet sizes. The Ariel space telescope will provide spectroscopic observations of a statistical sample of large gaseous exoplanets \citep{Tinetti2018}, but will have limited sensitivity for lower-mass exoplanets. Ground-based extremely large telescopes (ELTs), with diameters of 25 to 39 m, are expected to provide spectroscopic observations of a handful of nearby low-mass exoplanets. The facilities listed above are currently under construction. 

Observing a statistical sample of Earth-like exoplanets orbiting a variety of diverse host stars will require space-based survey facilities that are currently in their early planning stages, such as the Habitable Worlds Observatory (HWO) \citep{Stark2024} and Large Interferometer for Exoplanets (LIFE) \citep{Quanz2022}. These missions would provide observations of Earth- like planets around Sun-like stars that are analogous to the current exploration of M-dwarf rocky exoplanets with JWST, across optical to mid-infrared wavelengths. Insights gained from JWST observations will feed into the survey design of HWO and LIFE, including whether M-dwarf exoplanets should be a continued focus in the search of habitable conditions. Current exoplanet research therefore provides a path toward finding Earth-like exoplanets and determining their habitability.

\vspace{0.3cm}
\noindent \textbf{Acknowledgments:} 
T.L., J.T., and E.M.-R.K. acknowledge the Alfred P. Sloan Foundation’s AEThER project (G202114194) for providing the opportunity to discuss the ideas in this Review in depth.
\noindent \textbf{Funding:}
T.L. acknowledges support from the Branco Weiss Foundation, the Netherlands eScience Center under grant NLESC.OEC.2023.017, and NASA’s Nexus for Exoplanet System Science research coordination network (Alien Earths project, 80NSSC21K0593). O.S. acknowledges support from STFC grant UKRI1184.
\noindent \textbf{Author contributions:}
Conceptualization: T.L., O.S., J.T., E.M.-R.K. Data curation: T.L., J.T. Formal analysis: T.L., O.S., J.T., E.M.-R.K. Investigation: T.L., O.S., J.T., E.M.-R.K. Methodology: T.L., O.S., J.T. Visualization: T.L., O.S., J.T. Writing – original draft: T.L., O.S., J.T. Writing – review \& editing: T.L., O.S., J.T., E.M.-R.K.
\noindent \textbf{Competing interests:}
There are no competing interests to declare.
\noindent \textbf{Data and materials availability:}
No new data were generated for this Review.

\bibliography{refs1,refs2}{}
\bibliographystyle{aasjournal}

\end{document}